\documentstyle[12pt]{article}
\begin{document}
\baselineskip 20pt

\noindent
\hspace*{13.7cm}
hep-ph/9708316\\
 
\vspace{1.8cm}
 
\centerline{\Large \bf A Few Notes on Non-perturbative Parameters}
\vspace{0.5cm}

\centerline{\Large \bf in Heavy Quark Expansion}
 
\vspace{0.8cm}  
 
\begin{center}  
{{\bf Chao-shang~Huang}$^{a,c}$~~~and~~~{\bf C. S.~Kim}$^{b,c}$

\vspace{0.5cm}

{\it $a:$ Institute of Theoretical Physics, Academia Sinica, 100080
Beijing, China}\footnote{csh@itp.ac.cn}\\
{\it $b:$ Department of Physics, Yonsei University, 
Seoul 120-749, Korea}\footnote{kim@cskim.yonsei.ac.kr,~~
cskim@kekvax.kek.jp}\\
{\it $c:$ I.C.T.P., P.O. Box 586, 34100 Trieste, Italy}}
 
\vspace{1.0cm}
 
(\today)
\end{center}
\vspace{0.5cm}

\begin{abstract}

\noindent
Non-perturbative parameters to the order 
${\cal O}({1 \over m_Q})$ in HQET expansion,
$\overline{ \Lambda}$, $\lambda_1$ and $\lambda_2$,
as well as the masses, $m_b$ and $m_c$, are estimated
phenomenologically from $B$ and $D$ meson system spectroscopy.
We found relatively large value of $m_b$ 
and quite small value of $\overline{ \Lambda}$
compared to previous estimates.

\end{abstract}
\newpage

[{\bf I}]~~The heavy quark effective theory (HQET) provides a systematic
expansion in $1 \over m_Q$ to study physics of heavy hadrons containing
a heavy quark \cite{hqet}, and has been applied to various area of
phenomenology \cite{hqetph,hqetsl}. 
In $m_Q \rightarrow \infty$ limit the effective
theory has a spin--flavor symmetry (heavy quark symmetry). 
Applied
to the heavy hadron spectra, the symmetry leads to sum rules relating
masses of heavy hadrons with different heavy flavors. 
{\it E.g.} using the sum rules,
the masses of some beauty hadrons have been predicted in terms of the
masses of known charm hadrons \cite{9}.

The observed mass splitting of the ground-state spin doublet is
141 (or 46) MeV for $D$ (or $B$) mesons. 
It implies that the $1 \over m_Q$
corrections to the leading order are important for the realistic heavy
hadron spectra, especially, for charm hadrons. Furthermore, the
parameters of the order ${\cal O}({1 \over m_Q})$ 
represent the contributions from the
kinetic and chromo-magnetic operator \cite{10} and are the fundamental
parameters of HQET, which play a crucial role in the description of power
corrections to the leading order. 
For example, the leading power corrections
to $B \rightarrow X_q + e \bar \nu$ decay rate are completely
determined in terms of the parameters \cite{hqetsl}, 
which parameterize the nonperturbative effects. 
However, unfortunately, the parameters have not been 
calculated theoretically in an unambiguous way. 
Therefore, to estimate them from
phenomenological analyses is quite meaningful. 
In this short note we study the
spectra of heavy mesons to the order ${\cal O}({1 \over m_Q})$ 
in the framework of heavy quark expansion, and estimate the parameters of 
the order ${\cal O}({1 \over m_Q})$ from phenomenological analyses.

{}From the effective Lagrangian in HQET, the mass of a heavy hadron can
be written as
\begin{equation}
M_{h} = m_{Q} + \overline{ \Lambda} + \frac{1}{m_{Q}}\left(a 
+ < \vec s_{Q} \cdot \vec j_{l}> b\right),
\end{equation}
where $\overline{ \Lambda} $ is the ``mass'' of light degree of freedom
from the binding energy and masses of the light quarks. 
$a$ and $b$ are the parameters which characterize the effects
of the heavy flavor symmetry breaking and spin symmetry breaking at the
order ${\cal O}({1 \over m_Q})$, respectively:
\begin{eqnarray}
\overline{\Lambda} &=& \lim \limits_{m_Q \rightarrow \infty} (M_h - m_Q),
\nonumber\\
a &\equiv & -{1 \over 2} \lambda_1, \nonumber\\
b &\equiv &~~ 2 \lambda_2. \nonumber
\end{eqnarray}
Evidently $\overline{ \Lambda}$, $a$ and $b$ do not depend 
on heavy quark mass,
$m_Q$.
In Eq. (1) $\vec s_Q$ is the heavy quark spin, and $\vec j_l$ is the spin
of the light system. 
The measurements of $D$ and $B$ meson masses are now quite accurate, 
and from Eq. (1) we have
\begin{eqnarray}
 \overline{ M}_{D}& \equiv & \frac {1}{4}(M_{D}+3M_{D^*})=m_{c}+ 
\overline{ \Lambda} + \frac{a}{m_{c}} ~\simeq (1973 \pm 2)~{\rm MeV}, \\
\Delta M_{D}& \equiv & M_{D^*}- M_{D} = 
\frac{b}{m_{c}} ~\simeq (141 \pm 1)~{\rm MeV}, \\
\overline{ M}_{B} & \equiv & \frac {1}{4}(M_{B}+3M_{B^*}) = 
m_{b}+ \overline{ \Lambda} + \frac{a}{m_{b}} 
~\simeq (5313 \pm 2) ~{\rm MeV}, \\
\Delta M_{B}& \equiv & M_{B^*}- M_{B} = 
\frac{b}{m_{b}} ~\approx 46~{\rm MeV},
\end{eqnarray}
where we also show the experimental values from PDG96 \cite{PDG96}.

Eqs. (2-5) have five unknown independent quantities,
$(m_c,~m_b,~\overline{\Lambda},~a~{\rm and}~b)$, with four independent
observables. 
We choose $m_b$ as a free parameter\footnote{Here the parameter $m_b$ is 
phenomenologically defined only through Eq. (1) as an expansion parameter.
It might be a pole mass or a running mass or else 
depending on given extra assumptions.
In any case, however, it together with other parameters 
must satisfy Eq. (1) and the experimental constraints of Eqs. (2-5). }, 
so that we find the other parameters as functions of $m_b$:
\begin{eqnarray}
m_{c}& = & r \times m_{b} ~\simeq 0.326 \times m_b,\\
\overline{\Lambda} & = & \left(\overline{ M}_{D}+
\frac{\Delta \overline{ M}}{1-r}\right)-(1+r)\times m_{b}
~\simeq (6928 \pm 2)~{\rm MeV} - 1.326 \times m_b,\\
a & = & r \times m_{b} \cdot 
\left(m_{b}-\frac{\Delta \overline{ M}}{1-r}\right)
~\simeq 0.326 \times m_b \cdot 
\left(m_b - (4955 \pm 3)~{\rm MeV}\right),\\
b & = & \Delta M_{B} \times m_{b} ~\simeq 46~{\rm MeV} \times m_b,
\end{eqnarray}
where 
\begin{eqnarray}
r & \equiv & \frac{m_c}{m_b} =
\frac{\Delta M_{B}}{\Delta M_{D}} ~\simeq 0.326, \nonumber\\
\Delta\overline{ M} & \equiv & 
\overline{ M}_B-\overline{ M}_D ~\simeq (3340 \pm 2)~{\rm MeV}. \nonumber
\end{eqnarray} 
It is obvious that 
Eqs. (6-9) are not the definitions of the parameters, but
phenomenological relations satisfying the experimental 
results of Eqs. (2-4).

If we fix $m_b$=4800 MeV, which is deduced from a QCD analysis of the
$\Upsilon$ system, we get
\begin{displaymath}
   m_{c} \simeq 1565~{\rm MeV},~~ 
\overline{\Lambda} \simeq 563~{\rm MeV},~~
a \simeq -155~{\rm MeV}\times m_{c},~~
b \simeq 0.221~{\rm GeV}^2.
\nonumber
\end{displaymath}
Note that the value of $a$ becomes negative. This nagaive value for
$a$, certainly, can not be acceptable because the parameter $a$
is essentially the expectation value of 
the kinetic energy operator of the heavy quark 
in the rest frame of the hadron in HQET \cite{10,NEU}, 
and should be positive. 
This implies that the effective mass of 
the $b$ quark inside the $B$ meson is 
different, probably within order ${\cal O}(\Lambda _{QCD})$, 
from that inside the $\Upsilon$ system if
the expansions of heavy hadron masses in $1 \over m_Q$ are valid. 

In Fig. 1 we show 
the values of $m_c$ (in GeV), $\overline{ \Lambda}$ (in GeV), 
$a \equiv -{1 \over 2}\lambda_1$ (in GeV$^2$) and
$b \equiv 2 \lambda_2$ (in GeV$^2$) as functions of $m_b$.
As shown in Eq. (8), only for $m_b > 4958$ MeV the value of $a$ becomes
positive. Similary, only for $m_b < 5223$ MeV 
the value of $\overline{\Lambda}$ can be positive from Eq. (7). 
The shaded region, $4958 < m_b < 5223$ (in MeV), represents
for both $a>0$ and $\overline{\Lambda} >0$.
{}From the shaded region we can find the region of parameter space allowed
in HQET expansion to the order ${\cal O}({1 \over m_Q})$:
\begin{eqnarray}
m_b &=& [4958, ~5223]~~{\rm MeV}, \\
m_c &=& [1616, ~1703]~~{\rm MeV}~~~
({\rm or}~~(m_b-m_c)~=~[3342, ~3520]~~{\rm MeV}), \\
\overline{\Lambda} &=& [0, ~356]~~{\rm MeV}, \\
a &\equiv & -{1 \over 2} \lambda_1 ~=~ [0, ~0.461]~~{\rm GeV}^2, \\
b &\equiv &~~ 2 \lambda_2 ~~=~ [0.228, ~0.240]~~{\rm GeV}^2. 
\end{eqnarray}
\\

\begin{figure}[htb]
\mbox{}
\vskip20.cm\relax\noindent\hskip-2.0cm\relax
\vskip6cm\hskip-2cm
\includegraphics{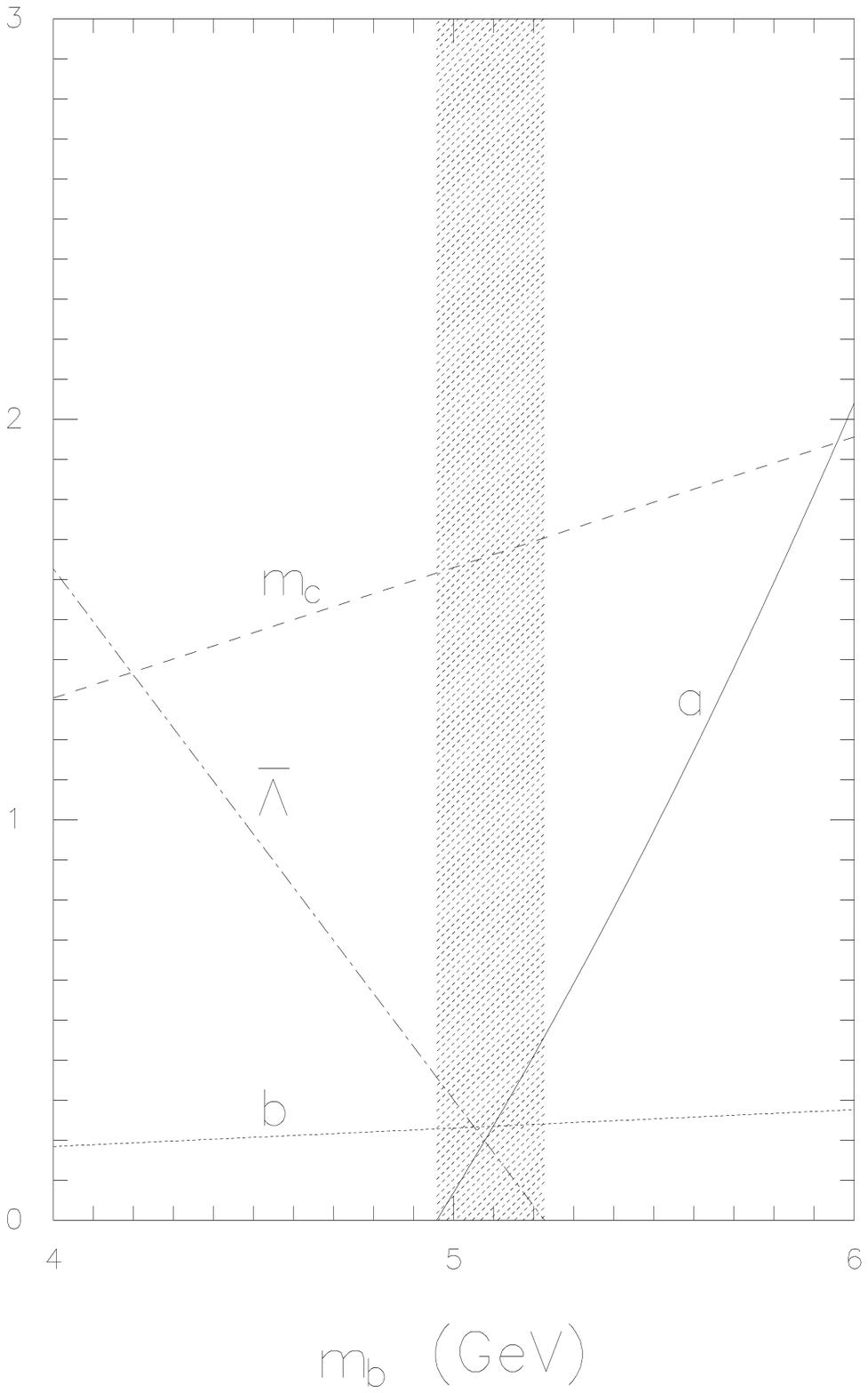} \vskip-6cm
\caption{\footnotesize 
The values of $m_c$ (in GeV), $\overline{ \Lambda}$ (in GeV),
$a \equiv -{1 \over 2}\lambda_1$ (in GeV$^2$) and
$b \equiv 2 \lambda_2$ (in GeV$^2$) as functions of $m_b$.
The shaded region, $4958 <m_b < 5223$ (in MeV), represents
for both $\overline{ \Lambda} >0$ (see Eq. (7)) and $a>0$ (see Eq. (8)).
}
\end{figure}

{\bf [II]}~~Previously the values of parameters have been estimated, 
{\it e.g.} see Ref. \cite{NEU}, as:
\begin{eqnarray}
m_b &= & (4.71 \pm 0.07)~~{\rm GeV}, \nonumber\\
m_b - m_c &= & (3.39 \pm 0.04)~~{\rm GeV}, \nonumber\\
\overline{\Lambda} &= & (0.57 \pm 0.07)~~
{\rm GeV},~~~({\rm see}~[8]) \nonumber\\
-\lambda_1 &= & (0.3 \pm 0.2)~~
{\rm GeV}^2,~~~({\rm see}~[9,10]) \nonumber\\
\lambda_2 &\approx & 0.12~~{\rm GeV}^2. \nonumber
\end{eqnarray}
Compared to our numerical bounds, Eqs. (10-14), previos estimations give
relatively small value of $m_b$ and quite large $\overline{\Lambda}$,
however at the same time, 
very compatible values for $(m_b-m_c)$, $\lambda_1$
and $\lambda_2$. As is well known, the heavy quark mass difference,
$(m_b-m_c)$, and spin symmetry breaking, $\lambda_2$, 
are precisely determined within 5 \% from $B$ and $D$ meson spectroscopy.
There are intrinsic uncertainties for the definitions of
$m_Q$, $\overline{\Lambda}$ and  $\lambda_1$ related to higher order
perturbative corrections \cite{NEU,int}. 
However, we note that Eqs. (1-9) are perfectly valid
independent of the definition of parameters in heavy quark mass
expansion up to ${\cal O}({1 \over m_Q})$. 
Therefore, the bounds on the
parameters, Eqs. (10-14), and the experimental constraints, Eqs. (2-5), 
have to be  
satisfied simultaneously as pre-requirements: {\it e.g.} a parameter set 
chosen from central values of the previous estimates
$$
(m_b,~m_c,~\overline{\Lambda},~-\lambda_1,~\lambda_2)=
(4.71~{\rm GeV},~1.32~{\rm GeV},~0.57~{\rm GeV},
~0.3~{\rm GeV}^2,~0.12~{\rm GeV}^2)
$$
can not simulataneously satisfy the experimental results, Eqs. (2-5),
and should be discarded as a heavy quark expansion parameter set,
even though an individual parameter can have value outside or inside 
the bound depending on its proper definition.
We also note that if the value of $m_b$ is chosen as $\sim 4.7$ GeV,
the only parameter set consistent with Eqs. (2-5) is given as
$$
(m_b,~m_c,~\overline{\Lambda},~-\lambda_1,~\lambda_2) \sim
(4.7~{\rm GeV},~1.5~{\rm GeV},~0.7~{\rm GeV},
~-0.8~{\rm GeV}^2,~0.11~{\rm GeV}^2),
$$
which does not satisfy the bounds at all.

Recently, it has been an important subject to obtain an accurate value 
of the kinetic energy,
$\mu_\pi^2~ (\equiv -\lambda_1)$, of the heavy quark inside $B$-meson.
Ball $et$ $al.$ \cite{theory} calculated using the QCD sum rule
approach and obtained $\mu_\pi^2 \sim 0.50 $ ${\rm GeV}^2$
for $B$-meson, while Neubert \cite{theory} obtained
$-\lambda_1 \sim 0.1$ ${\rm GeV}^2$.
It should be noted that those two derivations
differ in the choice of the 3-point correlation functions used to
estimate the matrix elements of interest.
Bigi $et$ $al.$ \cite{theory} derived an inequality between the 
expectation value of the kinetic energy operator of the heavy quark
inside the hadron and that of the chromomagnetic operator,
$\mu_\pi^2  \,\ge\, {3 \over 4} ({M_V}^2-{M_P}^2)$,
which gives $\mu_\pi^2 \ge 0.36\ {\rm GeV}^2$ for $B$-meson system.
However,
Kapustin $et$ $al.$ \cite{theory} showed later that this lower bound
could be significantly weakened 
by higher order perturbative corrections to the 3-point functions.
Hwang $et$ $al.$ \cite{theory} also calculated  the value 
by applying the variational method to the relativistic Hamiltonian,
and obtained $\mu_\pi^2 \sim 0.44$ GeV$^2$.
Similarly de Fazio \cite{theory} computed the matrix elements of the
kinetic energy operator by means of a QCD relativistic potential
model, and found $\mu_\pi^2 \sim 0.46$ GeV$^2$.
Besides the theoretical calculations of $\mu_\pi^2$,
Gremm $et$ $al.$ \cite{exp} extracted the average kinetic energy
by comparing the prediction of the HQET \cite{hqetsl} with
the shape of the inclusive $B \rightarrow X l {\nu}$
lepton energy spectrum, and obtained
$-\lambda_1 = 0.19 \pm 0.10\ {\rm GeV}^2$.

If we choose the previously estimated average $\lambda_1$ value
\cite{NEU,theory,exp}, $-\lambda_1=(0.3 \pm 0.2)$ GeV$^2$, 
we can find quite narrow region of parameter space from the given
experimental bounds, Eqs. (2-9), and Fig. 1: 
\begin{equation}
m_b = (5.05 \pm 0.07)~{\rm GeV},~~
(m_b-m_c) = (3.41 \pm 0.04)~{\rm GeV},~~
\overline{\Lambda} = (0.23 \pm 0.08)~{\rm GeV}.
\end{equation}
The experimental value of semileptonic decay width of $B$ meson can give
the value $m_b \eta_{QCD}^{1/5}$ \cite{alta}, 
with QCD correction $\eta_{QCD}$,
by comparing with theoretical inclusive semileptonic decay width. 
After using ${\cal B}_{sl}$ and $\tau_B$ from PDG96 \cite{PDG96}
and using $\eta_{QCD}=0.77 \pm 0.05$ \cite{QCD},
we can get $m_b=(5.1 \pm 0.2)$ GeV, which is in good agreement with 
our result.
It is interesting  to note that the value of hard pole mass $m_b$
obtained by Chernyak \cite{chern} is very close to that in Eq. (15).
Finally we note that higher order corrections of 
${\cal O}({1 \over m_Q^2})$ can change the bounds of our estimation.
However, as shown in  \cite{NEU,theory}, the correction to $m_b$ is about
$\sim \pm 0.004$ GeV, 
and we can safely conclude that the higher order effect
can not affect our results at all.\\

{\bf Acknowledgement}\\

CSH thanks ICTP, and CSK thanks ICTP in Italy and the
Japanese Government for their fellowship on visit to ICTP, 
where we completed the work. 
The work of CSH was supported in part by National
Natural Science Foundation of China.
The work of CSK was supported in part by the CTP  of SNU, in
part by the BSRI Program BSRI-97-2425, 
and in part by the KOSEF-DFG, Project No. 96-0702-01-01-2.



\begin{thebibliography}{20}

\bibitem{hqet} M.B. Voloshin and M.A. Shifman,  Sov. J. Nucl. Phys.
{\bf 47} (1988) 511; N. Isgur and M.B. Wise,  Phys. Lett. {\bf B232}
(1989) 113, {\bf B237} (1990) 527;
E. Eichten and B. Hill, Phys. Lett. {\bf B234} (1990) 511;
A.F. Falk, H. Georgi, B. Grinstein and M.B.Wise, 
Nucl. Phys. {\bf B343} (1990) 1.

\bibitem{hqetph} M.E. Luke, Phys. Lett. {\bf B252} (1990) 447;
T. Mannel, W. Roberts and Z. Ryzak,  Phys. Lett. {\bf B255}
(1991) 593; M. Neubert, Phys. Lett. {\bf B264} (1991) 455;
Z. Hioko {\it et al.},  Phys. Lett. {\bf B299} (1993) 115;
Y.B. Dai, X.H. Guo and C.S. Huang,  Nucl. Phys. {\bf B412}
(1994) 277.

\bibitem{hqetsl} 
J. Chay, H. Georgi and B. Grinstein, Phys. Lett. {\bf B247} (1990) 399; 
I.I. Bigi, M.A. Shifman, N.G. Uraltsev and   
A.I. Vainshtein, Phys. Rev. Lett. {\bf 71} (1993) 496;
A.V. Manohar and M.B. Wise, Phys. Rev. {\bf D49} (1994) 1310;
B. Blok, L. Koyrakh, M. Shifman and A.I. Vainshtein, 
Phys. Rev. {\bf D49} (1994) 3356; 
A. Falk, M. Luke and M. Savage, Phys. Rev. {\bf D49} (1994) 3367;
T. Mannel, Nucl. Phys. {\bf B423} (1994) 396.

\bibitem{9} V. Aglietti, Phys. Lett. {\bf B281} (1992) 341.

\bibitem{10} A.F. Falk and M. Neubert, Phys. Rev. {\bf D47}(1993) 2965;
{\bf D47} (1993) 2982.

\bibitem{PDG96} Particle Data Group 96: Phys. Rev. {\bf D54} (1996) 1.

\bibitem{NEU} M. Neubert, Phys. Rept. {\bf 245} (1994) 259;
hep-ph/9702375 (Feb. 1997). 

\bibitem{lamb} H.D. Politzer and M.B. Wise, 
Phys. Lett. {\bf B206} (1988) 681;
E. Bagan, P. Ball, V.M. Braun and H.G. Dosch, 
Phys. Lett. {\bf B278} (1992) 457.

\bibitem{theory} 
P. Ball and V.M. Braun, Phys. Rev. {\bf D49} (1994) 2472;
M. Neubert, Phys. Lett. {\bf B389} (1996) 727;
I.I. Bigi, M.A. Shifman, N.G. Uraltsev and
A.I. Vainshtein, Int. J. of Mod. Phys. {\bf A9} (1994) 2467;
Phys. Rev. {\bf D52} (1995) 196;
A. Kapustin, Z. Ligeti, M.B. Wise and B. Grinstein,
Phy. Lett. {\bf B375} (1996) 327;
D.S. Hwang, C.S. Kim and W. Namgung,
Z. Phys. {\bf C69} (1995) 107; Phys. Rev. {\bf D53} (1996) 4951;
Phys. Rev. {\bf D54} (1996) 5620; 
hep-ph/9608392, Phys. Lett. {\bf B} (in press);
F. de Fazio, Mod. Phys. Lett. {\bf A11} (1996) 2693.
 
\bibitem{exp} 
M. Gremm , A. Kapustin, Z. Ligeti and M.B. Wise,
Phys. Rev. Lett. {\bf 77} (1996) 20.

\bibitem{int} For example, see M. Neubert, Phys. Lett. {\bf B393}
(1997) 110; I. Bigi, M. Shifman and N. Uraltsev, hep-ph/9703290,
and References therein.
 
\bibitem{alta} G. Altarelli, G. Martinelli, S. Petraca and F. Rapuano,
Phys. Lett. {\bf B382} (1996) 409.

\bibitem{QCD} P. Ball, M. Beneke and V.M. Braun, 
Phys. Rev. {\bf D52} (1995) 3929.

\bibitem{chern} V. Chernyak, hep-ph/9407353 (July 1994).

\end{thebibliography}
\end{document}